\documentclass[preprint]{aastex6}

\newcommand{\casa}{Cassiopeia~A}

\newcommand{\myemail}{toshiki@astro.isas.jaxa.jp}

\begin{document}


\title{Multi-Year X-ray Variations of Iron-K and Continuum Emissions \\ in the Young Supernova Remnant Cassiopeia A}


\author{Toshiki Sato\altaffilmark{1,2}, Yoshitomo Maeda\altaffilmark{2},
   Aya Bamba\altaffilmark{3,4},
   Satoru Katsuda\altaffilmark{5},
   Yutaka Ohira\altaffilmark{3},
   Ryo Yamazaki\altaffilmark{3},
   Kuniaki Masai\altaffilmark{1},
   Hironori Matsumoto\altaffilmark{6},
   Makoto Sawada\altaffilmark{3},
   Yukikatsu Terada\altaffilmark{7},
   John P. Hughes\altaffilmark{8}
   and Manabu Ishida\altaffilmark{2}
   }

\altaffiltext{1}{Department of Physics, Tokyo Metropolitan University, 1-1 Minami-Osawa, Hachioji, Tokyo 192-0397}
\altaffiltext{2}{Department of High Energy Astrophysics, Institute of Space and Astronautical Science (ISAS),Japan Aerospace Exploration Agency (JAXA), 3-1-1 Yoshinodai, Sagamihara, 229-8510, Japan; \myemail}
\altaffiltext{3}{Department of Physics, The University of Tokyo, 7-3-1 Hongo, Bunkyo-ku, Tokyo 113-0033, Japan}
\altaffiltext{4}{Research Center for the Early Universe, School of Science, The University of Tokyo, 7-3-1 Hongo, Bunkyo-ku, Tokyo 113-0033, Japan}
\altaffiltext{5}{Department of Physics, Faculty of Science \& Engineering, Chuo University, 1-13-27 Kasuga, Bunkyo, Tokyo 112-8551, Japan}
\altaffiltext{6}{Nagoya University, Furo-cho, Chikusa-ku, Nagoya 464-8602, Japan}
\altaffiltext{7}{Saitama University, Shimo-Okubo 255, Sakura, Saitama 338-8570, Japan}
\altaffiltext{8}{Department of Physics and Astronomy, Rutgers University, 136 Frelinghuysen Road, Piscataway, NJ. 08854-8019, USA}



\begin{abstract}
We found simultaneous decrease of Fe-K line and 4.2-6~keV continuum
of Cassiopeia A with the monitoring data taken by Chandra in 2000--2013.
The flux change rates in the whole remnant are $-$0.65$\pm$0.02 \% yr$^{-1}$ in the
4.2--6.0 keV continuum and $-$0.6$\pm$0.1 \% yr$^{-1}$ in the Fe-K. In the
eastern region where the thermal emission is considered to dominate, the
variations show the largest values: $-$1.03$\pm$0.05 \% yr$^{-1}$ (4.2-6 keV
band) and $-$0.6$\pm$0.1 \% yr$^{-1}$ (Fe-K line). In this region,
the time evolution of the emission measure and the temperature have a decreasing trend.
This could be interpreted as the adiabatic cooling with the expansion
of $m = 0.66$. On the other hand, in the non-thermal emission dominated
regions, the variations of the 4.2--6 keV continuum show the smaller
rates: $-$0.60$\pm$0.04 \% yr$^{-1}$ in the southwestern region,
$-$0.46$\pm$0.05 \% yr$^{-1}$ in the inner region and $+$0.00$\pm$0.07 \% yr$^{-1}$
in the forward shock region. In particular, the flux does not show
significant change in the forward shock region. These results imply
that a strong braking in the shock velocity has not been occurring in
Cassiopeia A ($<$ 5 km s$^{-1}$ yr$^{-1}$). All of our results
support that the X-ray flux decay in the remnant is mainly caused by the
thermal components.
\end{abstract}


\section{Introduction}

Supernova remnants (SNRs) are known to be one of the most dynamic phenomenon in
the Universe. The spectral and image evolutions are quicker when the age
of the remnants are younger since the shock velocity is faster and its
braking is larger. Recently, there are several arguments about the X-ray
spectral variations from young SNRs
\citep*[e.g.,][]{2007Natur.449..576U,2008ApJ...677L.105U,2011ApJ...729L..28P}.
Mainly, the variable component is considered to be synchrotron X-rays
(non-thermal X-rays) caused by high-energy electrons in the amplified
magnetic field ($\sim$mG). Also, time series of images revealed
moments of expanding shell structures in young SNRs
\citep[e.g., ][]{2006ApJ...645..283F,2008ApJ...678L..35K,2009ApJ...697..535P}.
These facts tell us that SNRs are experiencing an extreme evolution, and
we can detect these evolutions in our observational time-scale ($\sim$10 yr).
Such an information would be very useful for understanding how the
remnants evolve and effect the ambient medium.

{\casa}, a Galactic young remnant of $\sim340$ yrs old \citep*{2006ApJ...645..283F},
has been found to display several X-ray time variations by intensive observations
with the {\it Chandra} observatory. \citet{2007AJ....133..147P} found year-scale
X-ray variability in thermal and non-thermal knots using the {\it Chandra} data
taken in 2000, 2002, and 2004. In the entire face of the remnant, they
identified six time-varying structures, four of which show count rate
increase from $\sim$10 \% to over 90 \%. \cite{2008ApJ...677L.105U} analyzed
the same dataset and found year-scale time variation in the X-ray intensity
for a number of non-thermal X-ray filaments or knots associated with the
reverse-shocked regions. They found that variable non-thermal features are
much more prevailing than the thermal ones. \citet{2011ApJ...729L..28P} found
a steady $\sim$1.5--2 \% yr$^{-1}$ decline in the 4.2--6.0 keV band of the
overall X-ray emission of {\casa}. They discussed a possible cause of this
decline as a deceleration of the forward shock velocity. The strong braking
with $\approx$30--70 km s$^{-1}$ yr$^{-1}$ was necessary to explain the decay
of the X-ray flux.

On the other hand, we cannot completely ignore a contribution of the thermal
X-rays to the time evolution because the continuum emissions below 4 keV have
not only the non-thermal component but also the thermal bremsstrahlung component.
\citet{2008ApJ...686.1094H} estimated that the fraction of non-thermal component
is $\sim54$ \% in the 4.2--6 keV band. In the further lower energy band (1-3 keV),
no significant X-ray variability in the soft X-ray band (1--3 keV) has been found
\citep*[e.g.,][]{2011ApJ...729L..28P,2013ApJ...769...64R}. However, we note the possibility
that the 4.2--6 keV and the 1--3 keV band emission may originate
from different plasma. The spectrum of {\casa} can be well fitted with two
temperature thermal model \citep*[e.g.,][]{2002A&A...381.1039W,2009ApJ...703..883H}
in addition to the non-thermal component. Of the two thermal components, the higher
temperature one occupies a significant fraction of the observed 4.2--6 keV spectrum,
and explains the entire Fe-K line. In addition, spatial distribution of the iron in
Cassiopeia A is not similar to hard X-ray intensity distribution
\citep*[e.g.,][]{2015ApJ...802...15G}. The thermal component has a different spatial
distribution from the non-thermal component, and hence we believe it is worth while
to investigate time variation of the thermal component.

In this paper, while placing a possibility of time variation of the thermal component
in mind, we aimed to identify the variable component of the 4.2--6 keV of {\casa} in
more detail. We investigated the time variations in the 4.2--7.3 keV band including
the Fe-K emission lines by using {\it Chandra} ACIS for the first time. The time
variations in thermal emission dominated and non-thermal emission dominated regions
were also investigated.

\section{Observation and Data Reduction}

\subsection{Chandra ACIS-S}
For our study of the year-scale variability in flux, data of {\it Chandra X-ray Observatory}
were utilized. The {\it Chandra}  observations of {\casa} have been carried out several
times since the launch in 1999
\citep*[e.g.,][]{2000ApJ...537L.119H,2004ApJ...615L.117H,2011ApJ...729L..28P,2012ApJ...746..130H,2014ApJ...789..138P}.
The data used in our analysis are listed in Table \ref{tab:chandradata}. The archived data
taken with ACIS-S with a Timed Exposure(TE) mode are gathered. ACIS-S3 is the back-illuminated CCD
chip with enhanced soft X-ray response and fairly constant spectral resolution during the course of the mission
compared to the ACIS-I array. The satellite and instrument are
described by \cite{2002PASP..114....1W}.

We reprocessed the event files (from level 1 to level 2) to remove pixel randomization and
to correct for CCD charge transfer efficiencies using CIAO version 4.6 and CalDB 4.6.3.
The bad grades were filtered out and good time intervals were reserved. {\casa} is so
bright that the data were all taken with the single chip operation mode of S3 to avoid
the telemetry loss. Cassiopeia A was usually pointed near the center of the ACIS-S3 chip
of 1024 $\times$ 1024 pixel CCDs, each with $0^{\prime\prime}.5 \times 0^{\prime\prime}.5$
pixels and a field of view of $8'.4 \times 8'.4$. The pointing position is moderately
shifted from the aim-point and its roll angle varies from observation to observation.
The effective areas (arf) of individual observations were then calculated for each ObsID
using the {\it Chandra} standard analysis software package mkwarf in CIAO.

\begin{table}
\caption{{\it Chandra} observation log.}
\begin{center}
\small
\begin{tabular}{cccc}
\hline
ObsID. &Date&Exposure (ks)&SI mode\\
&YYYY/MM/DD&&\\ \hline
114 .& 2000/01/30&49.9&TE\_002A0\\
1952 .& 2002/02/06&49.6&TE\_002A0\\
5196 .& 2004/02/08&50.2&TE\_002A0\\
5319 .& 2004/04/18&42.3&TE\_003DA\\
9117 .& 2007/12/05&24.8&TE\_003DA\\
9773 .& 2007/12/08&24.8&TE\_003DA\\
10935 .& 2009/11/02&23.3&TE\_009E2\\
12020 .& 2009/11/03&22.4&TE\_009E2\\
10936 .& 2010/10/31&32.2&TE\_009E2\\
13177 .& 2010/11/02&17.2&TE\_009E2\\
14229 .& 2012/05/15&49.1&TE\_009E2\\
14480 .& 2013/05/20&48.8&TE\_009E2\\\hline
\end{tabular}
\label{tab:chandradata}
\end{center}
\end{table}

\subsection{Suzaku XIS0 \& XIS3}

For tracing the non-thermal emission, the  {\it Suzaku}  data were utilized. A deep
observation with {\it Suzaku} was made in 2012. The exposure time was 205 ksec long
(XIS0+XIS3). {\it Suzaku} has four X-ray CCD cameras
\citep*[XIS: ][]{2007PASJ...59S..23K, 2007SPIE.6686E..0PU}. One of the four XIS
detector (XIS 1) is back-side illuminated (BI) and the other three (XIS 0, XIS 2 and XIS 3)
are front side illuminated (FI). In the XIS data taken with the Spaced-row Charge Injection
(SCI) option with the normal exposure mode, the gap columns due to the injected charges
appear at every 54 lines. The column widths of the FIs are three pixels which are smaller
than five for the BI. To minimize the flux uncertainty due to the gap, we used only
the FI data. In the FI CCDs, the XIS-2 were not operated in 2012. Data screening
was made with the standard criteria provided by the {\it Suzaku} processing team.

\section{Analysis and Results}

\begin{figure}[h]
 \begin{center}
  \includegraphics[width=8cm]{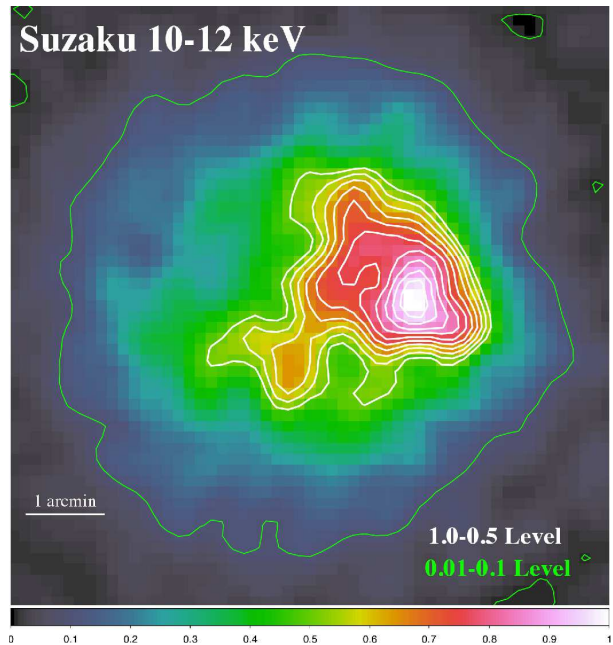}
 \end{center}
\caption{{\it Suzaku} XIS image which was corrected for the exposure map (i.e, the
vignetting function of the telescope) in the 10--12 keV band. This image was binned by
8 $\times$ 8 pixels, and was then smoothed with a Gaussian function with a sigma
of 3 bins (= 0.42 arcmin). The image is normalized by the value of the pixel
value of the maximum brightness.
White and green contours show the range of 1.0(maximum brightness)--0.5  and 0.1--0.01, respectively.}
\label{fig:Cas_region}
\end{figure}

\subsection{Region Selection}
\label{sec:selection}

X-ray emissions from {\casa} are known to be a composite of the thermal and non-thermal
components \citep*[e.g.,][]{2000ApJ...528L.109H}. The thermal component is characterized
by emission lines from highly ionized ions of heavy elements such as iron, accompanied
with a continuum emission by a thermal bremsstrahlung. The non-thermal emission is known
to be traced by a hard band continuum emission above 10 keV \citep[e.g.,][]{2009PASJ...61.1217M}.
Figure~\ref{fig:Cas_region} shows the hard X-ray image (10--12 keV band) with {\it Suzaku}.
We can see the concentrated hard X-ray distribution from the western to region toward
the center of Cassiopeia A. Figure \ref{fig:each_vari} shows the three color images of
{\it Chandra} in the 4.2--6.0 keV, the 6.54--6.92 keV (Fe-K line) and the 1.75--1.95
keV (Si-K line) bands overlaid with the {\it Suzaku} contour in the 10-12 keV band.
The Fe-K emission is believed to originate from the optically thin thermal plasma.
Using the Suzaku contour map and the Chandra lines, we can segregate the distributions
of the thermal and non-thermal X-rays. Based on this information, five local regions
and one whole SNR region were selected from the image for our analysis.

We defined the ``East" and the``North West" regions as the ``Thermal dominant" region
(Magenta ellipses in Figure~\ref{fig:each_vari}). The East region has the most abundant
X-ray flux of the Fe-K line. Therefore, this region is the best region
to discuss the time evolution of the Fe-K line with less contribution of non-thermal
emission. The North West region is the second luminous region of the Fe-K
line emission. This region shows a separation from the hard X-ray peak and a separation
from continuum X-ray dominant region \citep*{2004ApJ...613..343D,2008ApJ...686.1094H}.

On the other hand, we defined ``South West", ``Inner" and ``Forward Shock" region as
``Non-thermal dominant" region (Light blue regions in Figure~\ref{fig:each_vari}).
\citet{2008ApJ...686.1094H} and \citet{2004ApJ...613..343D} show the area dominated
by a harder spectrum. As a matter of fact, these regions are manifest themselves in hard
X-rays. Our image also shows the Inner region is bright with hard X-ray. In addition,
the forward shock has a featureless non-thermal emission
\citep*[e.g.,][]{2000ApJ...528L.109H,2005ApJ...621..793B}. In the forward shock,
it was found that the average proper motion of Cassiopeia A is $0.30^{\prime\prime}$ yr$^{-1}$
\citep*{2009ApJ...697..535P}, so we corrected the Forward Shock region of each year
with this value (see Appendix).

\subsection{Spectra}

\begin{figure*}[h]
 \begin{center}
  \includegraphics[width=16cm]{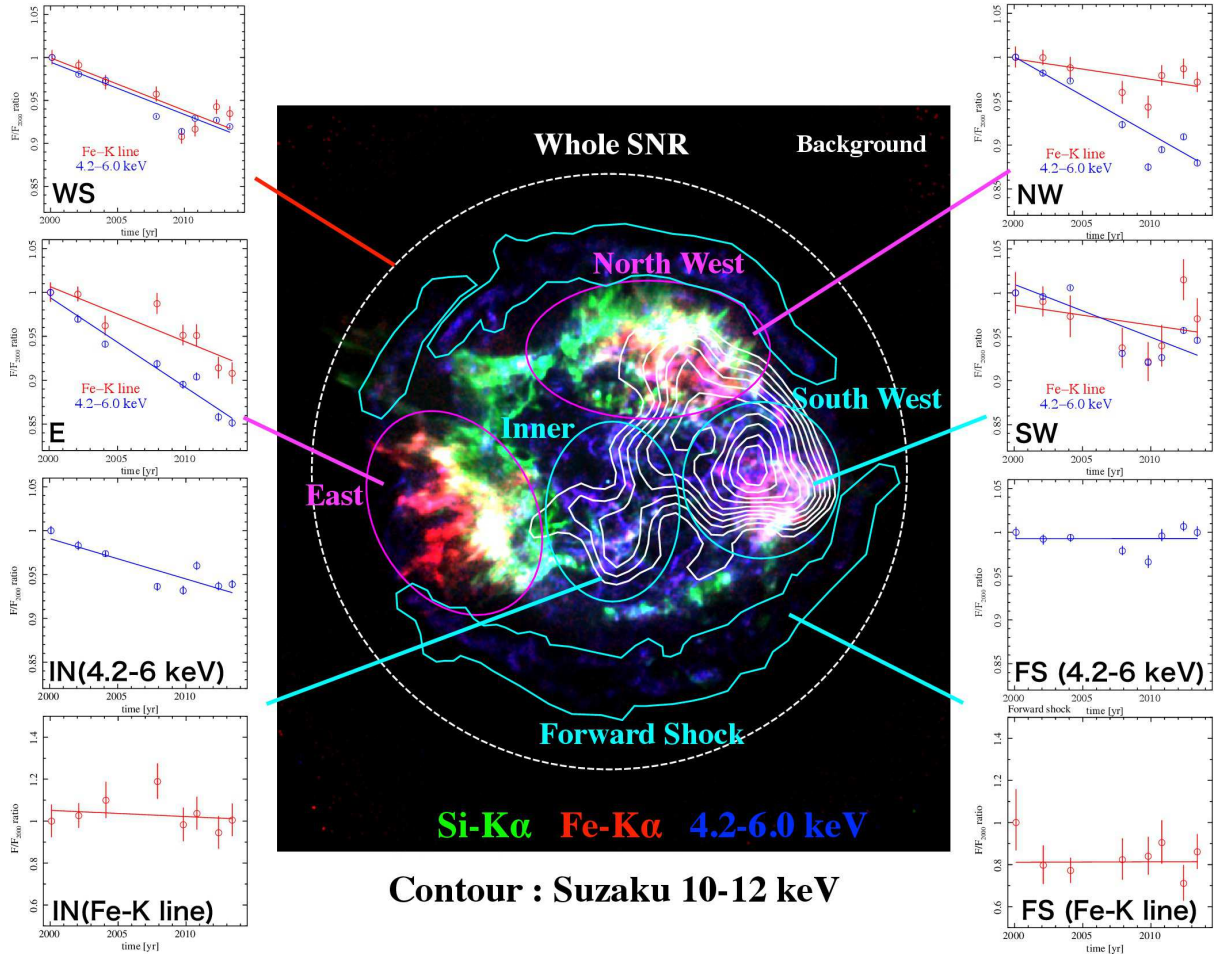}
 \end{center}
\caption{Three color images of {\casa} with {\it Chandra} overlaid with the {\it Suzaku} contour
in the 10-12 keV band. Blue, red and green colors show the 4.2--6.0 keV, 6.54--6.92 keV (Fe-K line)
and 1.75--1.95 keV (Si-K line) band images, respectively. Each plot around the image shows the
time evolution of observed flux of Cassiopeia A at each region (Whole SNR: WS, East: E, Inner: I,
North West: NW, South West: SW, Forward Shock: FS). Blue and red data show flux of the 4.2--6 keV
band and Fe-K line normalized at the first data-point, respectively. Solid lines show the best-fit
linear models. The error bars of all Figures are 1 $\sigma$.}
\label{fig:each_vari}
\end{figure*}

With the regions defined in the section \ref{sec:selection}, we extracted the spectra using
a custom pipeline based on \verb"specextract" in CIAO script. Here, the background spectra
were extracted from outside of the Whole SNR region defined in Figure~\ref{fig:each_vari}.
Since the X-ray emission from Cassiopeia A is very strong, the background contribution is
almost negligible for an estimation of the time variation. The background fraction of the whole
SNR is only $\sim$3 \% in 4.2--7.3 keV band and this is almost constant value from 2002 to
2013. In 2000, this fraction shows a larger value ($\sim$7 \%) presumably because of an increase
in the charged particle flux experienced during this observation. After the background subtraction,
we fitted the 4.2-7.3 keV band spectra
at each epoch with a power-law model and a Gaussian line with XSPEC version 12.8.2. The best-fit
parameters are summarized in Table \ref{tab:fit_results}.

\begin{figure*}[h]
 \begin{center}
  \includegraphics[width=16cm]{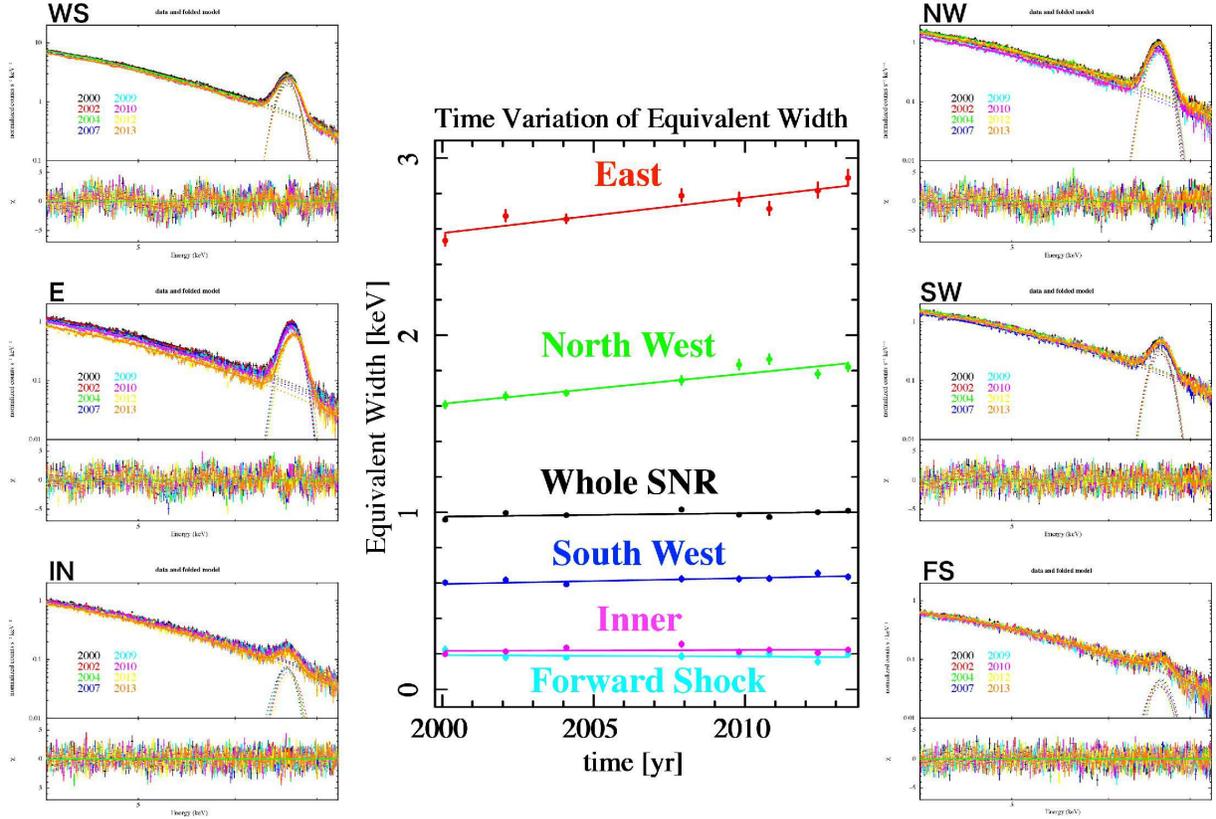}
 \end{center}
\caption{Time variation of equivalent width of Fe-K line (plot in the middle) and observed spectra
in 4.2-7.3 keV band (on both sides) at each region (Whole SNR: WS, East: E, Inner: I, North West:
NW, South West: SW, Forward Shock: FS). In the central plot, the solid lines show the best-fit
linear models. Individual spectra were fitted with a model composed of a power law and a Gaussian.
The error bars of all Figures are 1 $\sigma$.}
\label{fig:spec_EW}
\end{figure*}

Figure~\ref{fig:spec_EW} shows the spectra in the 4.2--7.3 keV band taken from the six regions.
As described in the section \ref{sec:selection}, the two bright non-thermal dominant regions
South West and Inner are remarkable in the hard X-ray continuum flux (Figure~\ref{fig:Cas_region}
and \ref{fig:spec_EW}). Accordingly, the photon indexes of these regions are $\sim$2.6--3.0 while
those of the thermal are 3.0--3.4. In the Forward Shock and the Whole SNR, the photon indexes and
that evolutions are a slightly different from the results in \cite*{2011ApJ...729L..28P}.
This is probably due to difference of the energy band used in the spectral fitting. The fit residuals
are larger for the thermal dominant than for the non-thermal regions. This is because weak thermal
lines such as Cr-K (5.6 keV) appeared in the band.

As shown in Figure~\ref{fig:spec_EW} and Table \ref{tab:decayrate}, we also found the increase of
the equivalent width of the Fe-K line of Cassiopeia A for the first time. In the Thermal dominant
regions, we can see $\sim 10$ \% increasing of the equivalent width for 13 years, and it is notable
that a large evolution of the equivalent width like this is the first detection from all the
supernova remnants. In addition, we found the Fe-K line centroid varies among the observations.
Since, there is a calibration uncertainty for Fe K-line centroids of $\sim$0.3 \% (or $\sim$20
eV at Fe-K)\footnote{Available at {\tt http://cxc.harvard.edu/cal/docs/cal\_present\_status.html} \\}
, however, it is difficult to discuss its evolution.

\vspace{0.5cm}

\begin{table*}
\tiny
\begin{center}
\caption{Best-fit parameters of the ACIS spectra\tablenotemark{a}}
\label{tab:fit_results}
\end{center}
\begin{center}
\begin{tabular}{cccccccc}
\hline
		  	& 	 Epoch	& 			\multicolumn{2}{c}{power-law model}				&								\multicolumn{3}{c}{Fe-K line}						&$\chi^2$/d.o.f \\
	 		& 		 	&	$\Gamma$ 				& 	Flux (4.2--6 keV) 		&			 Energy 		&		 EW 						&	Flux 						&				 \\
			& [yr] 		& 							& [${\times}10^{-11}$ erg cm$^{-2}$ s$^{-1}$] &	 			[keV]		&		[keV]				&	[${\times}10^{-3}$ ph cm$^{-2}$ s$^{-2}$] &\\\hline
\multicolumn{2}{l}{Whole SNR}  	& & &\\
			&2000.1 	& 	2.90$\pm$0.02 			& 	 17.74$^{+0.02}_{-0.04}$ 	& 	 6.6287$^{+0.0012}_{-0.0008}$ 			& 	0.95$\pm$0.01 			& 	5.27$\pm$0.05			& 	347.52/206	\\
			&2002.1 	& 	2.99$\pm$0.02 			& 	 17.39$\pm$0.03 			& 	 6.6309$^{+0.0013}_{-0.0007}$ 			& 	1.00$\pm$0.01 			&	5.22$\pm$0.04			& 	322.02/206 	\\
			&2004.1 	& 	2.99$\pm$0.01 			& 	 17.27$^{+0.01}_{-0.03}$ 	& 	 6.6376$^{+0.0010}_{-0.0005}$ 			& 	0.98$\pm$0.01 			&	5.12$\pm$0.03			& 	355.18/206	\\
			&2007.9  	& 	3.00$\pm$0.02 			& 	 16.52$\pm$0.03 			& 	 6.6377$\pm$0.001 						& 	1.02$\pm$0.01 	 		&	5.05$\pm$0.04			& 	324.81/206 	\\
			&2009.8 	& 	3.00$\pm$0.02 			& 	 16.21$\pm$0.03 			& 	 6.6423$^{+0.0014}_{-0.0009}$ 			& 	0.99$\pm$0.01 			&	4.79$\pm$0.04			& 	321.93/206	\\
			&2010.8 	& 	2.99$\pm$0.02 			& 	 16.48$\pm$0.03 			& 	 6.6354$^{+0.0014}_{-0.0008}$ 			& 	0.97$\pm$0.01 			&	4.83$\pm$0.04			& 	322.59/206	\\
			&2012.4 	& 	2.95$\pm$0.02 			& 	 16.44$\pm$0.03 			& 	 6.6503$\pm$0.001			 			& 	1.00$\pm$0.01 	 		&	4.97$\pm$0.04			& 	278.20/206	\\
			&2013.4 	& 	3.00$\pm$0.02 			& 	 16.31$\pm$0.03 			& 	 6.6419$\pm$0.001			 			& 	1.01$\pm$0.01 	 		&	4.93$\pm$0.04			& 	268.78/206	\\
\multicolumn{2}{l}{East} & & &\\
			&2000.1 	& 	3.22$\pm$0.04 			& 	2.52$\pm$0.01 				& 	6.677$^{+0.002}_{-0.001}$				& 	2.53$\pm$0.03 			& 	1.77$\pm$0.02			& 	291.35/206 	\\
			&2002.1 	& 	3.31$\pm$0.04 			& 	2.45$\pm$0.01 				& 	6.675$^{+0.002}_{-0.001}$ 				& 	2.67$^{+0.04}_{-0.03}$ 	& 	1.76$\pm$0.02			& 	307.28/206 	\\
			&2004.1 	& 	3.29$\pm$0.03 			& 	2.37$\pm$0.01 				& 	6.685$\pm$0.002				 			& 	2.66$^{+0.03}_{-0.02}$ 	& 	1.70$\pm$0.01			& 	420.74/206 	\\
			&2007.9 	& 	3.30$\pm$0.04 			& 	2.32$\pm$0.01 				& 	6.677$^{+0.002}_{-0.001}$	 			& 	2.79$^{+0.04}_{-0.03}$ 	& 	1.75$\pm$0.02			& 	299.12/206 	\\
			&2009.8 	& 	3.31$\pm$0.04 			& 	2.26$\pm$0.01 				& 	6.679$^{+0.002}_{-0.001}$	 			& 	2.76$\pm$0.04 			& 	1.68$\pm$0.02			& 	278.59/206 	\\
			&2010.8 	& 	3.28$\pm$0.04 			& 	2.28$\pm$0.01 				& 	6.672$^{+0.002}_{-0.001}$	 			& 	2.71$^{+0.04}_{-0.03}$ 	& 	1.68$\pm$0.02			& 	270.84/206 	\\
			&2012.4 	& 	3.30$\pm$0.05 			& 	2.17$\pm$0.01 				& 	6.711$^{+0.002}_{-0.001}$	 			& 	2.82$^{+0.05}_{-0.04}$ 	& 	1.62$\pm$0.02			& 	307.66/206 	\\
			&2013.4 	& 	3.40$\pm$0.05 			& 	2.15$\pm$0.01 				& 	6.704$^{+0.002}_{-0.001}$	 			& 	2.89$^{+0.05}_{-0.04}$ 	& 	1.61$\pm$0.02			& 	292.35/206 	\\
\multicolumn{2}{l}{North West} & & &\\
			&2000.1 	& 	2.98$\pm$0.03 	 		& 	3.73$^{+0.01}_{-0.02}$		& 	6.584$^{+0.002}_{-0.001}$	 			& 	1.61$\pm$0.02 			& 	1.85$\pm$0.02			& 	316.05/206 	\\
			&2002.1 	& 	3.03$\pm$0.03 			& 	3.66$\pm$0.01 				& 	6.589$^{+0.002}_{-0.001}$ 				& 	1.66$\pm$0.02 			& 	1.85$\pm$0.02			& 	283.29/206 	\\
			&2004.1 	& 	3.06$\pm$0.02 			& 	3.63$\pm$0.01 				& 	6.597$\pm$0.001				 			& 	1.67$\pm$0.02 			& 	1.83$\pm$0.02			& 	389.43/206 	\\
			&2007.9 	& 	3.12$\pm$0.03 			& 	3.44$\pm$0.01 				& 	6.596$^{+0.002}_{-0.001}$	 			& 	1.74$^{+0.03}_{-0.02}$ 	& 	1.78$\pm$0.02			& 	277.33/206 	\\
			&2009.8 	& 	3.15$\pm$0.04 			& 	3.26$^{+0.01}_{-0.02}$ 		& 	6.605$^{+0.002}_{-0.001}$	 			& 	1.83$\pm$0.03 			& 	1.75$\pm$0.02			& 	287.39/206 	\\
			&2010.8 	& 	3.17$\pm$0.04 			& 	3.34$\pm$0.01 				& 	6.597$^{+0.002}_{-0.001}$	 			& 	1.86$\pm$0.03 			& 	1.81$\pm$0.02			& 	232.13/206 	\\
			&2012.4 	& 	3.01$\pm$0.03 			& 	3.39$\pm$0.01 				& 	6.618$^{+0.002}_{-0.001}$	 			& 	1.78$\pm$0.02 			& 	1.83$\pm$0.02			& 	267.84/206 	\\
			&2013.4 	& 	3.04$\pm$0.03 			& 	3.28$^{+0.02}_{-0.01}$ 		& 	6.604$^{+0.002}_{-0.001}$	 			& 	1.82$\pm$0.02 			& 	1.80$\pm$0.02			& 	290.10/206 	\\
\multicolumn{2}{l}{South West} & & &\\
			&2000.1 	& 	2.74$\pm$0.03 			& 	3.92$^{+0.02}_{-0.01}$ 		& 	6.614$^{+0.003}_{-0.002}$	 			& 	0.60$\pm$0.01 			& 	0.77$\pm$0.02			& 	242.97/206 	\\
			&2002.1 	& 	2.83$\pm$0.03 			& 	3.90$\pm$0.01 				& 	6.623$^{+0.003}_{-0.002}$ 				& 	0.62$^{+0.02}_{-0.01}$ 	& 	0.76$\pm$0.02			& 	214.20/206 	\\
			&2004.1 	& 	2.78$\pm$0.02 			& 	3.94$\pm$0.01 				& 	6.630$\pm$0.002				 			& 	0.59$\pm$0.01 			& 	0.75$\pm$0.01			& 	229.44/206 	\\
			&2007.9 	& 	2.82$\pm$0.03 			& 	3.65$^{+0.01}_{-0.02}$ 		& 	6.627$\pm$0.003				 			& 	0.62$\pm$0.02 			& 	0.72$\pm$0.02			& 	265.07/206 	\\
			&2009.8 	& 	2.82$\pm$0.03 			& 	3.61$^{+0.02}_{-0.01}$ 		& 	6.631$\pm$0.003				 			& 	0.62$\pm$0.02 			& 	0.71$\pm$0.02			& 	220.24/206 	\\
			&2010.8 	& 	2.80$\pm$0.03 			& 	3.63$\pm$0.01 				& 	6.627$^{+0.003}_{-0.002}$	 			& 	0.62$^{+0.02}_{-0.01}$ 	& 	0.73$\pm$0.02			& 	265.88/206 	\\
			&2012.4 	& 	2.80$\pm$0.04 			& 	3.75$\pm$0.01 				& 	6.636$^{+0.003}_{-0.002}$	 			& 	0.66$\pm$0.02 			& 	0.78$\pm$0.03			& 	230.55/206	\\
			&2013.4 	& 	2.82$\pm$0.03 			& 	3.71$\pm$0.01 				& 	6.629$^{+0.003}_{-0.002}$	 			& 	0.64$\pm$0.02 			& 	0.75$\pm$0.02			& 	264.19/206 	\\
\multicolumn{2}{l}{Forward Shock} & & &\\
			&2000.1 	& 	2.55$\pm$0.06 			& 	1.57$\pm$0.01 				& 	6.63$^{+0.02}_{-0.01}$				 	& 	0.22$\pm$0.02			& 	0.12$\pm$0.02 			& 	198.16/206 	\\
			&2002.1 	& 	2.64$\pm$0.05 			& 	1.562$^{+0.008}_{-0.009}$ 	& 	6.61$^{+0.02}_{-0.01}$					& 	0.18$\pm$0.02 			& 	0.10$\pm$0.01 			& 	208.52/206 	\\
			&2004.1 	& 	2.69$\pm$0.04 			& 	1.565$^{+0.005}_{-0.007}$ 	& 	6.627$^{+0.008}_{-0.009}$		 		& 	0.18$\pm$0.01 			& 	0.092$\pm$0.007 		& 	199.11/206 	\\
			&2007.9 	& 	2.61$\pm$0.05 	 		& 	1.541$^{+0.009}_{-0.007}$ 	& 	6.64$\pm$0.02				 			& 	0.19$\pm$0.02 			& 	0.10$\pm$0.01 			& 	229.06/206 	\\
			&2009.8 	& 	2.67$\pm$0.05 	 		& 	1.52$\pm$0.01 				& 	6.62$\pm$0.01				 			& 	0.20$\pm$0.02 			& 	0.10$\pm$0.01 			& 	192.79/206 	\\
			&2010.8 	& 	2.55$\pm$0.05 	 		& 	1.57$\pm$0.01 				& 	6.63$\pm$0.01				 			& 	0.20$\pm$0.02 			& 	0.11$\pm$0.01 			&	207.04/206 	\\
			&2012.4 	& 	2.57$\pm$0.05 	 		& 	1.585$^{+0.009}_{-0.008}$ 	& 	6.61$^{+0.02}_{-0.01}$		 			& 	0.16$\pm$0.02 			& 	0.08$\pm$0.01 			& 	193.40/206 	\\
			&2013.4 	& 	2.65$\pm$0.05 	 		& 	1.575$^{+0.010}_{-0.008}$ 	& 	6.64$\pm$0.01				 			& 	0.20$\pm$0.02 			& 	0.10$\pm$0.01 			& 	205.78/206 	\\
\multicolumn{2}{l}{Inner} & & &\\
			&2000.1 	& 	2.86$\pm$0.04 	 		& 	2.409$^{+0.012}_{-0.009}$ 	& 	6.612$^{+0.008}_{-0.007}$	 			& 	0.20$\pm$0.01 			& 	0.15$\pm$0.01		 	& 	188.96/206 	\\
			&2002.1 	& 	2.93$\pm$0.04 			& 	2.37$\pm$0.01 				& 	6.631$^{+0.009}_{-0.008}$	 			& 	0.21$\pm$0.01 			& 	0.15$\pm$0.01		 	& 	203.19/206 	\\
			&2004.1 	& 	3.01$\pm$0.03 			& 	2.346$^{+0.006}_{-0.008}$ 	& 	6.625$^{+0.006}_{-0.007}$	 			& 	0.23$\pm$0.01 			& 	0.17$\pm$0.01		 	& 	236.49/206 	\\
			&2007.9 	& 	2.89$\pm$0.04 			& 	2.256$^{+0.008}_{-0.011}$ 	& 	6.627$^{+0.009}_{-0.008}$	 			& 	0.26$\pm$0.02 			& 	0.18$\pm$0.01		 	& 	187.30/206 	\\
			&2009.8 	& 	2.88$\pm$0.04 			& 	2.24$\pm$0.01 				& 	6.618$\pm$0.009				 			& 	0.21$\pm$0.01 			& 	0.15$\pm$0.01			& 	181.55/206 	\\
			&2010.8 	& 	2.93$\pm$0.04 			& 	2.31$\pm$0.01 				& 	6.649$^{+0.010}_{-0.009}$	 			& 	0.22$\pm$0.02 			& 	0.16$\pm$0.01		 	& 	230.67/206 	\\
			&2012.4 	& 	2.91$\pm$0.04 			& 	2.26$\pm$0.01 				& 	6.655$^{+0.010}_{-0.009}$	 			& 	0.21$\pm$0.02 			& 	0.14$\pm$0.01		 	& 	215.17/206 	\\
			&2013.4 	& 	2.99$\pm$0.04 			& 	2.26$\pm$0.01 				& 	6.632$\pm$0.009				 			& 	0.22$\pm$0.02 			& 	0.15$\pm$0.01 			& 	179.23/206 	\\
\hline
\end{tabular}
\end{center}
\begin{scriptsize}
\tablenotetext{1}{The errors are at 1 $\sigma$ confidence level. }
\end{scriptsize}
\end{table*}

\subsection{Time Variation of 4.2-6.0 keV and Fe-K}

\begin{table}[h]
\begin{center}
\caption{Time variation of Cassiopeia A in 4.2-7.3 keV band\tablenotemark{a}.}\label{tab:decayrate}
\begin{tabular}{llcccc}
\hline
& 					&	  4.2--6.0 keV 			&	      Fe-K 				&		Equivalent Width 		\\
& 					&	[\% yr$^{-1}$] 			&		[\% yr$^{-1}$]		&		[\% yr$^{-1}$] 		\\ \hline
\multicolumn{2}{l}{Whole SNR} 		&	$- 0.65 \pm 0.02 $ 	&	$- 0.6 \pm 0.1 $  		&	$+ 0.2 \pm 0.1$ 		\\ \hline
\multicolumn{3}{l}{Thermal dominant region} &														\\
& East 				&	$- 1.03 \pm 0.05 $ 	&	$- 0.6 \pm 0.1 $  		&	$+ 0.8 \pm 0.2$ 		\\
& North West 		&	$- 0.88 \pm 0.04 $ 	&	$- 0.2 \pm 0.1 $  		&	$+ 1.1 \pm 0.2$ 		\\ \hline
\multicolumn{3}{l}{Non-thermal dominant region} &													\\
& South West 		&	$- 0.60 \pm 0.04 $ 	&	$- 0.2 \pm 0.3 $  		&	$+ 0.6 \pm 0.3$ 		\\
& Inner 			&	$- 0.46 \pm 0.05 $ 	&	$- 0.3 \pm 0.9 $		&	$+ 0.3 \pm 0.9$ 		\\
& Forward Shock 	&	$+ 0.00 \pm 0.07 $ 	&	$- 0.0 \pm 1.2 $		&	$- 0.3 \pm 1.0$ 		\\
\hline
\end{tabular}
\end{center}
\begin{scriptsize}
\tablenotetext{1}{The errors are at 90 \% confidence level. }
\end{scriptsize}
\end{table}

\begin{figure}[h]
 \begin{center}
  \includegraphics[width=16cm]{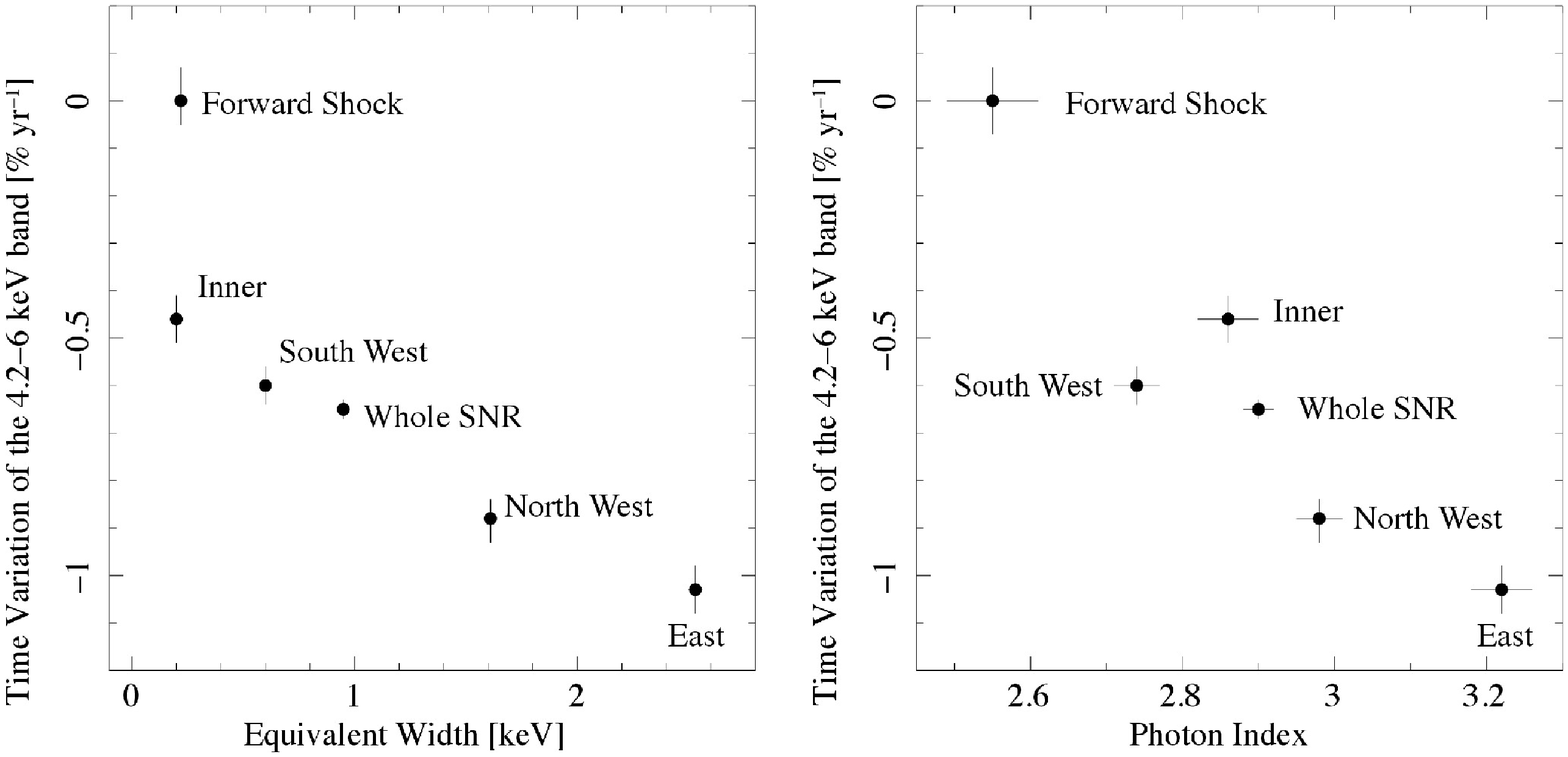}
 \end{center}
\caption{Left: a plot of the change rate of the 4.2-6 keV continuum intensity versus the equivalent
width of the Fe-K line. Right: a plot of the change rate of the 4.2-6 keV continuum flux versus the
photon index of the continuum power law. The equivalent width and the photon index are the best fit
values in 2000 yr (Table \ref{tab:fit_results}). The error bars are 90 \% confidence level.}
\label{fig:vari_EW}
\end{figure}

From the fitting, we investigated the time variation of the 4.2-6.0 keV band and Fe-K line fluxes.
Flux evolution of the {\casa} was well reproduced by linear decline \citep*{2011ApJ...729L..28P}.
Therefore, we also fitted time variation of the fluxes with a linear model. These fitting results
of each region is summarized in Figure~\ref{fig:each_vari} and Table \ref{tab:decayrate}. In the
Whole SNR, the 4.2--6 keV band and the Fe-K line fluxes show a significant decline in these
$\sim$10 years with a similar change ($\sim -0.6$ \% yr$^{-1}$).

There is a discrepancy between the variation of the 4.2-6 keV band in this work: $-0.65 \pm 0.02$ \% yr$^{-1}$
and that in \cite{2011ApJ...729L..28P}: $-1.5 \pm 0.17$ \% yr$^{-1}$, in spite of nearly the same data set.
Although we tried several analysis methods (see Appendix), we could not reveal the cause of the discrepancy.

In the local regions, time variation of the 4.2-6.0 keV continuum and Fe-K line fluxes are different
from the Whole SNR. In Figure~\ref{fig:vari_EW}, we can see larger time variation of the 4.2-6 keV
band in the regions which have higher equivalent width of the Fe-K line and the softer photon index.
In addition, we also found the Forward Shock region has no significant change of 4.2-6 keV and Fe-K.
From these results, we can interpret that Cassiopeia A is undergoing a flux change in the reverse
shock region which has a larger contribution of the Fe-K line.

\subsection{Fitting the spectra of the East region with the bremsstrahlung model}
\label{sec:bremss}

\begin{table}[h]
\begin{center}
\caption{Best-fit parameters of the bremsstrahlung model in the East region\tablenotemark{a}}\label{tab:fit_bremss}
\end{center}
\begin{center}
\begin{tabular}{cccccccc}
\hline
		  	& 	 Epoch& 			\multicolumn{2}{c}{bremss model}								                  & $\chi^2$/d.o.f 	\\
	 		& 		 	&		  kT  					   	& 		normalization\tablenotemark{b} 				&	   				\\
			& [yr] 	& 		[keV]					    	   &  														   &		    			\\\hline
			&2000.1 	& 	2.80$^{+0.10}_{-0.09}$ 			& 				8.65$^{+0.46}_{-0.43}$ 					& 	312.82/206 		\\
			&2002.1 	& 	2.69$^{+0.09}_{-0.08}$ 			& 				8.95$^{+0.48}_{-0.45}$ 					& 	307.28/206 		\\
			&2004.1 	& 	2.70$^{+0.07}_{-0.06}$ 			& 				8.62$^{+0.35}_{-0.34}$ 					& 	420.74/206 		\\
			&2007.9 	& 	2.68$\pm$0.09 				   	& 				8.51$^{+0.48}_{-0.45}$ 					& 	299.12/206 		\\
			&2009.8 	& 	2.66$^{+0.10}_{-0.09}$ 			& 				8.41$^{+0.51}_{-0.48}$ 					& 	278.59/206 		\\
			&2010.8 	& 	2.71$^{+0.10}_{-0.09}$ 			& 				8.25$^{+0.48}_{-0.45}$ 					& 	270.84/206 		\\
			&2012.4 	& 	2.67$\pm$0.10 					   & 				8.00$^{+0.51}_{-0.48}$ 					& 	307.66/206 		\\
			&2013.4 	& 	2.54$^{+0.10}_{-0.09}$  		& 				8.63$^{+0.56}_{-0.53}$ 					& 	292.35/206 		\\
\hline
\end{tabular}
\end{center}
\begin{scriptsize}
\tablenotetext{1}{The errors are at 1 $\sigma$ confidence level. }
\tablenotetext{2}{$3.02 \times 10^{-17}/4 {\pi} D^{2} \int n_e n_i dV$. }
\end{scriptsize}
\end{table}

As shown in Figure \ref{fig:vari_EW}, we found that the East region has the largest decay rate and
the largest equivalent width of the Fe-K line. Since the Fe-K line is a tracer of the thermal plasma
emission, we evaluated time variation of the emission from the East region via a thermal model.
Table \ref{tab:fit_bremss} shows the results of fitting with a thermal bremsstrahlung instead
of the power-law model. We then drew time histories of the resultant temperature and normalization
(=emission measure), and fitted them with a linear model. As a result, it is found that the time
evolution of the temperature and the emission measure are $-(0.4\pm0.3)$ \% yr$^{-1}$ and
$-(0.5\pm0.5)$ \% yr$^{-1}$ (90 \% confidence level), respectively.

\section{Discussion}

\citet{2011ApJ...729L..28P} already reported the decay of the intensity of the 4.2-6 keV continuum
from the whole remnant of {\casa}. They discussed the cause of the decay by assuming all the emission
is originated from the non-thermal emission. As shown in Figure~\ref{fig:each_vari} we found time
variation in the whole remnant is also observed at the Fe-K line. Moreover, if we pick-up the local
regions, the variabilities in the continuum and the Fe-K line are highly different from region to
region. Therefore, the cause of the time variation in the 4.2-6 keV continuum and the Fe-K lines
must be revisited. We positively use the fact that the Fe-K line is evidence of thermal emission.
Then, we found the Forward Shock region which has faint Fe-K emission shows the smallest decay rate
and the East region which has bright Fe-K emission shows the largest decay rate. The result naturally
supports that these regions have different variable components (thermal or non-thermal origin).
Using these regions, we here discuss the origin of the decay with the thermal dominant scenario
(section \ref{thermal_decay}) and non-thermal dominant scenario (section \ref{nonthermal_decay})
individually.

\subsection{A decay scenario of the thermal components}\label{thermal_decay}

Young SNRs like Cassiopeia A are experiencing a drastic expansion due to the high speed of their
ejecta. The plasma formed by the shock heating should be then adiabatically expanded. The adiabatic
expansion causes the cooling of the plasma, too. Therefore, the changing of X-ray flux from thermal
component must be first examined with an adiabatic cooling. \cite{2003ApJ...597..347L} suggested
that Cassiopeia A is currently transitioning from the ejecta-dominated to the Sedov--Taylor phase,
and hence it is natural to assume that Cassiopeia A is experiencing an adiabatic evolution
without a radiative cooling.

Here, we evaluate the time decay of the regions where the thermal emission is dominant by assuming
that their entire emission is of purely thermal origin. Bremsstrahlung X-ray flux from a thermal
plasma is described as
\begin{equation}
 F_\nu \propto EM \cdot T^{-\frac{1}{2}} {\rm exp} \left(-\frac{h\nu}{k_BT}\right) \cdot \bar{g}_{ff} \label{eq:themal}
\end{equation}
where $EM$, $h$, $k_B$ and $\bar{g}_{ff}$ are the emission measure, the Planck constant, the Boltzmann
constant and the Gaunt factor, respectively. The Gaunt factor is given by \citep[see][]{1979rpa..book.....R},
\begin{equation}
 \bar{g}_{ff} = \left( \frac{3}{\pi} \frac{k_BT}{h\nu} \right)^{\frac{1}{2}} ~~ {\rm for ~~}\frac{k_BT}{h\nu}<1, \label{eq:gaunt}
\end{equation}
\begin{equation}
 \bar{g}_{ff} = \frac{\sqrt{3}}{\pi} {\rm ln}\left( \frac{4}{\zeta} \frac{k_BT}{h\nu} \right) ~~ {\rm for ~~}\frac{k_BT}{h\nu}>1,
\end{equation}
where the constant $\zeta = 1.781$. In the case of Cassiopeia A, a typical electron temperature
is in the rage of 1-3 keV \citep*{2012ApJ...746..130H}. Therefore, we can use equation (\ref{eq:gaunt})
because the energy band we chose (4.2-6.0 keV) is above the spectrum cut-off.

First, we calculated time evolution of the emission measure. We assume the number of particle $nV$ = constant.
Using the expansion parameter: $m$ ($r \propto t^m$), we can describe $V \propto r^3 \propto t^{3m}$,
$n \propto V^{-1} \propto t^{-3m}$ around dynamical time scale or in the  beginning of the Sedov-Taylor
phase, and then
\begin{equation}
 EM = n^2 V \propto t^{-3m}
\end{equation}
where we assumed $n_e \sim n_i$. In the case of Cassiopeia A, the change rate of emission measure is
\begin{equation}
 \frac{1}{EM} \frac{dEM}{dt} = \frac{-3m}{t} \Rightarrow -0.58 \left(\frac{m}{0.66}\right) \left(\frac{t}{340}\right)^{-1} ~ {\rm ~ \% ~ yr^{-1}}. \label{em}
\end{equation}
where we normalized $m$ by 0.66 \citep*{2009ApJ...697..535P} and $t$ by the remnant age of 340 yr.

Next, we calculate thermal decay rate by taking adiabatic cooling into account.
For the adiabatic gas, $PV^{\gamma} = ~{\rm const.}$
By the same token, we can estimate temperature evolution along with the plasma volume as described below,
\begin{equation}
 TV^{\gamma - 1} = ~{\rm const.} ~ \Rightarrow ~ T \propto V^{1-\gamma} \propto t^{-2m}
\end{equation}
where $\gamma = 5/3$ is the heat capacity ratio. Thus, we can estimate the rate of decline of the
temperature is
\begin{equation}
 \frac{1}{T} \frac{dT}{dt} = \frac{-2m}{t} \Rightarrow -0.39 \left(\frac{m}{0.66}\right) \left(\frac{t}{340~\rm{yr}}\right)^{-1} ~ {\rm ~ \% ~ yr^{-1}}. \label{kT}
\end{equation}
Here we can describe $EM(t)=EM_0(t/t_0)^{-3m}$ and $T(t)=T_0(t/t_0)^{-2m}$. In order to calculate the
flux evolution including both of these effects, $F_\nu \propto EM(t) ~ T(t)^{-1/2} ~ {\rm exp}(-h\nu/k_B T(t)) ~ (k_BT(t)/h\nu)^{1/2}$,
and then
\begin{equation}
 \frac{dF_\nu}{dt} = -F_\nu \left( \frac{3m}{t} + \frac{2m}{t} \frac{h\nu}{k_B T} \right).
\end{equation}
We measured a typical electron temperature of Cassiopeia A is $k_BT \sim 2.5$ keV. This is about twice
as low as the mean photon energy of the 4.2-6.0 keV band. Thus we can estimate flux change rate of thermal
component as below,
\begin{equation}
 \frac{1}{F_\nu}\frac{dF_\nu}{dt} = -\frac{7m}{t} \Rightarrow -1.36 \left(\frac{m}{0.66}\right) \left(\frac{t}{340~\rm{yr}}\right)^{-1} ~ {\rm ~ \% ~ yr^{-1}}.\label{decay}
\end{equation}
The variation in the East region ($= -1.03$; see Table \ref{tab:decayrate}) is closest to this.
Also, the predicted rates of the emission measure and temperature (see eq.(\ref{em}) and eq.(\ref{kT}))
are consistent with the observational values in the East region: $-(0.4\pm0.3)$ \% yr$^{-1}$ for
the emission measure and $-(0.5\pm0.5)$ \% yr$^{-1}$ for the temperature. Thus we conclude that the
flux variation in the East region of Cassiopeia A could be explained by the thermal variation due to
the adiabatic expansion.

On the other hand, the change rates in the other regions are much smaller than the value predicted by eq.(\ref{decay}).
In particular, the variations in the non-thermal dominant regions ($< -0.6$ \% yr$^{-1}$) could not be
explained by the adiabatic expansion.

\subsection{A decay scenario of the non-thermal components}\label{nonthermal_decay}

The cosmic-ray electrons are considered to be accelerated in the shock front by diffusive shock acceleration
\citep*[DSA;][]{1978MNRAS.182..443B,1987PhR...154....1B}. In the Sedov phase, the blast wave is decelerated
by sweeping up ambient interstellar matter. This effect causes a flux decay of synchrotron X-rays. In addition
to this, evolution of the magnetic field and the electron injection rate also changes the flux of synchrotron
X-rays. Here, we investigated whether the evolution of these parameters could explain the time variation of
Cassiopeia A or not.

\begin{figure}[h]
 \begin{center}
  \includegraphics[width=12cm]{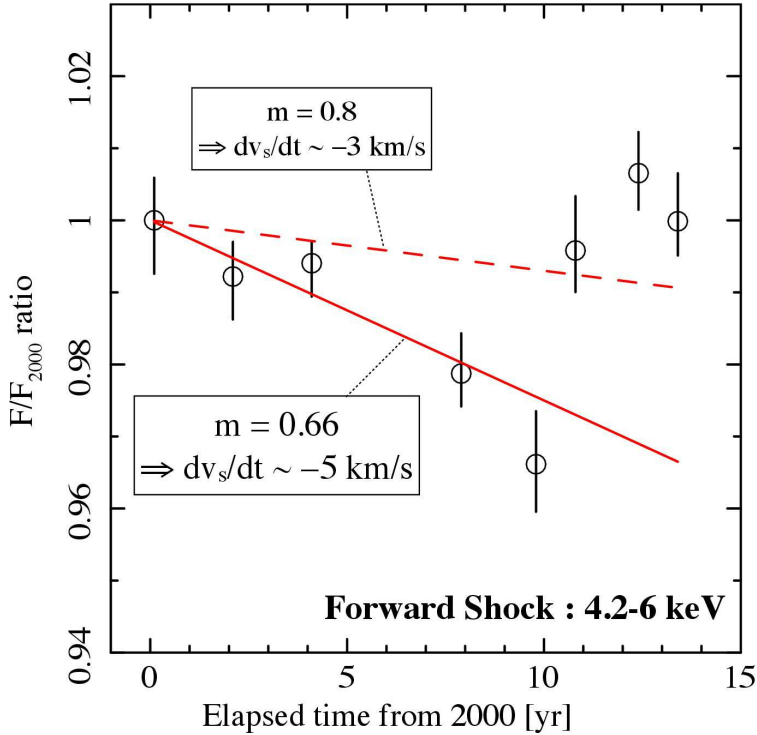}
 \end{center}
\caption{Comparison between the time observed variation and the predicted change rate in the forward shock
region in the band 4.2-6 keV. Black circles show our results of time variation in the Forward Shock (FS)
region. Red solid and broken lines show the predicted change rate with $m = 0.66$ and $m = 0.8$,
respectively.}
\label{fig:comp_model}
\end{figure}

The X-ray synchrotron emission is well approximated analytically. The energy spectrum of electrons is generally
given by
\begin{equation}
 N_e(E) = AE^{-p}(1 + E/E_b)^{-1}{\rm exp}[-(E/E_e^{\rm max})^2]
\end{equation}
where $A$, $E_b$ and $E_e^{\rm max}$ are a normalization factor, the break energy and the maximum energy of
electrons, respectively. In the case of Cassiopeia A, the radio index $\alpha = (p-1)/2 = 0.77$:
\cite{1977A&A....61...99B}. The break energy expresses the spectrum shape which suffers the synchrotron cooling.
During the acceleration, electrons with $E > E_b$ are losing their energies via synchrotron cooling, which provides
a steepened energy spectrum. From this electron distribution, we can calculate the approximate formula of the
X-ray luminosity as shown in the equation (5) of \cite{2012ApJ...746..134N},
\begin{equation}
L_{\nu} \propto AB_d^{(p+1)/2} {\nu}_b^{-(p-1)/2} (\nu/\nu_b)^{-p/2} {\rm exp}(- \sqrt{\nu/\nu_{\rm roll}}) \\ \nonumber \vspace{0.1cm}
\end{equation}
\begin{equation}
  \propto AB_d^{(p-2)/2} {\nu}^{-p/2} {\rm exp}(- \sqrt{\nu/\nu_{\rm roll}}), \hspace{+1.05cm} \label{eq11}
\end{equation}
where $B_d$, $\nu_b$ and ${\nu}_{\rm roll}$ are the downstream magnetic field, the break frequency and the
roll-off frequency, respectively. In this calculation, \cite{2012ApJ...746..134N} assumed the photon energy
is larger than the break photon energy ($\nu > \nu_b$), and the break frequency depends on the downstream
magnetic field ($\nu_b \propto B_d^{-3}$). This assumption could be adapted for young SNRs ($t_{\rm age} \lesssim 10^3$ yr)
due to their amplified magnetic field. Here, we attempted to estimate the time variation of the synchrotron X-ray
in the case of Cassiopeia A by transforming this equation into our framework of time evolution formulation.

First, we considered the time evolution of the normalization factor: $A$ in eq.(\ref{eq11}). We assumed
that the amount of accelerated particles is proportional to the product of the fluid ram pressure and the SNR
volume as assumed in \cite{2012ApJ...746..134N}. Then, we can describe
$A \propto (\rho v_s^2)r^3 \propto t^{3m-2}$, where $v_s$ is the shock velocity and we assumed the shock is moving
through the progenitor wind of the supernova: $\rho \propto r^{-2}$. Then the decay of this term is sensitive to the value of $m$.
In the case of Cassiopeia A ($m = 0.66$), we can find that this normalization is almost constant with time.

Second, we considered the time evolution of the term $B_d^{(p-2)/2}$ in eq.(\ref{eq11}).
The magnetic energy density is amplified to a constant fraction of $\rho v_s^2$: case (a) or $\rho v_s^3$:
case (a$^{\prime}$) \citep[e.g.,][]{2004MNRAS.353..550B,2006ESASP.604..319V,2008AIPC.1085..169V}, and we
can interpret the magnetic energy density evolution as a function of time below,
\begin{equation}
 B_d^2 \propto \rho v_s^2 \propto \frac{1}{r^2} \left(\frac{dr}{dt} \right)^2 \propto t^{-2} \Rightarrow B_d \propto t^{-1}, \label{B1}\\
\end{equation}
\begin{equation}
 B_d^2 \propto \rho v_s^3 \propto \frac{1}{r^2} \left(\frac{dr}{dt} \right)^3 \propto t^{(m-3)} \Rightarrow B_d \propto t^{\frac{1}{2}(m-3)}. \label{B2}
\end{equation}
Hereafter we denote $B_d^{(p-2)/2} \propto t^{X}$. If we {neglect the time evolution of $\nu_{\rm roll}$ for the time being},
the discussion so far results in the synchrotron intensity at the forward shock as $L_\nu \propto t^{X}$. Thus, we can
estimate the time variation of the synchrotron radiation by the evolution of the magnetic field as $1/L_\nu \cdot dL_\nu/dt = X \cdot t^{-1}$. In the case of
Cassiopeia A, $m = 0.66$: \cite{2009ApJ...697..535P}, $\alpha = 0.77$: \cite{1977A&A....61...99B} and $t_{\rm age} = 340$
yr predicts the variation of $-0.08$ \% yr$^{-1}$ for the case (\ref{B1}) and $-0.09$ \% yr$^{-1}$ for the case (\ref{B2}),
whose difference is quite small. From this result, we found that the contribution of the magnetic field evolution
to the X-ray variation is small.

Finally, we considered the time evolution of the term ${\nu}^{-p/2} {\rm exp}(- \sqrt{\nu/\nu_{\rm roll}})$
in eq.(\ref{eq11}). Assuming that $\nu_{\rm roll}$ is determined by a balance between the acceleration rate and the synchrotron
loss \citep{1999A&A...351..330A,2006MNRAS.371.1975Y}, we obtain below.
\begin{equation}
E^{\rm max}_e \propto B_d^{-1/2}v_s \Rightarrow \nu_{\rm roll} \propto {(E_e^{\rm max})}^2 B_d \propto v_s^2 \propto t^{2m-2} \equiv t^{Y}\\
\end{equation}
From the discussion of the time evolution all parameters ($A$, $B_d$ and $\nu_{\rm roll}$), the logarithmic derivative of eq.(\ref{eq11}),
results in \citep*[see also ][]{2010ApJ...723..383K},
\begin{equation}
 \frac{dL_\nu}{dt} = L_\nu \left(\frac{p}{t} + \frac{Y}{2t} \sqrt{\frac{\nu}{\nu_{\rm roll}}}\right) ~~; p = X+3m-2\\
\end{equation}
In the case of Cassiopeia A, roll-off energy $h\nu_{\rm roll}$ is suggested to be $\sim$ 2.3 keV in outer shock filament
\citep{2015ApJ...802...15G}. This implies $(\nu/\nu_{\rm roll}) \simeq 2$ for the band 4.2-6 keV. Thus, we can estimate
the variation in outer filament as $1/L_\nu \cdot dL_\nu/dt = 1/t \cdot (p + Y/\sqrt{2})$, and then a change rate is about
$-0.26$ \% yr$^{-1}$ for the both cases of (a) and (a$^{\prime}$).

Consequently the flux variability is sensitive to the value of the expansion parameter $m$. If we adopt $m = 0.66$, the variation in
the band 4.2-6 keV is estimated to be $-0.26$ \% yr$^{-1}$.
Figure \ref{fig:comp_model} shows a comparison between the predicted rate and the observed time variation of the 4.2-6 keV continuum in the
forward shock region. We found the model rate well fits to the observations from 2000 to 2010. The parameter $m$ could be interpreted as
a deceleration of the shock
velocity. If we assumed 5,000 km s$^{-1}$ as the shock velocity of Cassiopeia A, $m = 0.66$ means the deceleration of
$\sim 5$ km s$^{-1}$ yr$^{-1}$. In Fig. \ref{fig:comp_model}, however, the data points after 2010 do not follow the $m = 0.66$
line. For reference, we draw another line of $m = 0.8$ that is closer to the data after 2010. In this case  the
deceleration is $\sim 3$ km s$^{-1}$ yr$^{-1}$. This means a strong braking in the shock velocity has not been occurring in
the Forward Shock region (at most $\sim 5$ km s$^{-1}$ yr$^{-1}$). We can see a flux jump between 2010 and 2012. Several
non-thermal filaments in Cassiopeia A have a flickering of X-ray flux with a time scale of $\sim$year, and this jump might
also be able to be explained by that feature.

The particle acceleration and the synchrotron cooling in the reverse shock are very complicated. The continuum emission seems
to be decreasing gradually in the South West and the Inner regions. However, these regions have a number of flickering
filaments and have a large contribution of the thermal X-ray. In addition, the dynamical evolution of the reverse shock inward the
remnants have not been well understood. \cite{2014ApJ...785..130Z} investigated the particle acceleration in the forward and the
reverse shock of Cassiopeia A by using the numerical calculations. They predicted the change rate of the synchrotron X-ray in the
reverse shock is $\sim -0.9$ \% yr$^{-1}$ at least because the ejecta density drops proportional to $t^{-3}$ ( = $-$0.9 \% yr$^{-1}$).
However, we cannot see such a large decreasing in the South West and the Inner regions, and cannot explain the details of the
acceleration in the reverse shock in the present.

\section{Conclusion}

Our work shows the flux of the Fe-K in Cassiopeia A is decreasing with the continuum emissions in 4.2-6 keV for the first time.
By using hard X-ray distribution above 10 keV as a good indicator of non-thermal emission, we separated ``Thermal dominant" and
``Non-thermal dominant" regions from the whole SNR, and investigated their time variations. Then, we found clear correlations
of the decay rates in 4.2-6 keV band with the photon indexes and with the equivalent width of Fe-K line. The correlation shows
that the flux in the regions which have softer spectrum and richer emissions of Fe-K are decreasing more drastically.

We found the East region, which is considered to be ``Thermal dominant" region and has the softest spectrum ($\Gamma \sim 3.2$), shows
the most rapid decline. The flux change rate of the Fe-K line and 4.2-6 keV continuum are $-0.6\pm0.1$ \% yr$^{-1}$ and $-1.03\pm 0.05$ \% yr$^{-1}$,
respectively. In the region, the time evolution of continuum flux, emission measure and temperature are well explained by the
adiabatic cooling with the expansion of $r \propto t^{-m}$ with $m = 0.66$.

On the other hand, ``Non-thermal dominant'' regions show smaller decay rates. In particular, the Forward Shock region,
which has the hardest spectrum ($\Gamma \sim 2.6$), shows no large decay. It implies that the blast wave of  Cassiopeia A
does not seem to experience a strong deceleration \citep[such as $\approx$30--70 km s$^{-1}$ yr$^{-1}$:][]{2011ApJ...729L..28P}.
From the decay rate, we conclude that the deceleration is $\sim$ 5 km s$^{-1}$ yr$^{-1}$ at most.

It is interesting to note that the time evolution of the East region and the Forward Shock region, where the thermal emission
and the non-thermal emission dominates the most among all selected regions, respectively, can be represented by the power law
expansion of $r \propto t^{-m}$ with a common index of $m = 0.66$. The emission from the other regions is a certain mixture of
the thermal and non-thermal emission. Even though $m = 0.66$ is common, the resulting intensity decay rate is larger for the
thermal emission, and the intensity of the non-thermal continuum is nearly constant, if $m$ does not change in the last couple
of decade. A different mixing ratio probably results in a decay rate of the emission that is different from region to region.
Accordingly, we conclude that the decay of the X-ray intensity above $\sim$4 keV of the whole remnant is probably caused
by the thermal emission component.

\acknowledgments{T.S. is grateful for the travel support from HAYAKAWA FOUNDATION.
  This work was supported by the Japan Society for the Promotion of Science (JSPS)
  KAKENHI Grant Number 16J03448, 15K05107, 15K17657, 15K05088, 25105516 and 23540280.
  We thank Jacco Vink, Takayuki Hayashi and Ryo Iizuka for helpful discussion
  and suggestions in preparing this paper. We thank the anonymous referee for
  his/her comments that helped us to improve the manuscript.
}

\appendix

\section{Difference of Arf and source region}

We found a discrepancy of X-ray flux between our results and \cite{2011ApJ...729L..28P}. Table \ref{tab:diff} shows a comparison
of our results with other results which are analyzed with different methods. In this table, we calculated X-ray flux and count-rate
in the 4.2-6 keV band with several arfs and source regions for investigation of a cause of this discrepancy.

In CIAO, we can calculate two kinds of arfs (weighted arf for extended sources or imaging arf for pointlike source). We checked
whether the types of arfs have influence on an estimation of X-ray flux or not. The second and third columns in Table \ref{tab:diff}
show the 4.2-6 keV fluxes calculated by a weighted arf and an imaging arf, respectively. In this comparison, we cannot find a
large difference, and cannot see a large decay from 2000 to 2010 like Patnaude et al. (see the fifth column). Therefore, a difference
of arf type is not likely to be the cause of inconsistency. The fourth columns in Table \ref{tab:diff} show the fluxes calculated by an
imaging arf in a different source region (r = 2.5$'$). The region within 2.5$'$ circle in Cassiopeia A do not include a part of
the forward shock filaments (see Figure \ref{fig:region_def} left). In this case, whole flux shows a less value than 3.5$'$ circle
region, however the flux decay from 2000 to 2010 does not change that much. In the count rate, we can see a larger decay than flux decay,
as listed in the 6th and 7th columns of Table 5. This is because the effective area are decreasing with time. And, the decay of count rate
within 2.5$'$ circle is very similar to
\cite{2011ApJ...729L..28P}.

\begin{table*}
\scriptsize
\begin{center}
\caption{Difference of flux and count-rate of whole SNR in the 4.2-6 keV\tablenotemark{a}}
\label{tab:diff}
\end{center}
\begin{center}
\small
\begin{tabular}{cc|cccc|cc}
\hline
&								& 									&				\multicolumn{2}{c}{flux\tablenotemark{b}}								&									&	 				\multicolumn{2}{c}{count rate\tablenotemark{c}}		\\
&		Epoch					& 		arf for diffuse source		&				\multicolumn{2}{c}{arf for point source}					&			Patnaude et al.			&	 							&	 							\\
&		[yr] 					& 			r = 3.5$'$				& 			r = 3.5$'$				&		r = 2.5$'$						&									&		r = 3.5$'$				&		r = 2.5$'$				\\ \hline
&		2000					&		17.74$^{+0.02}_{-0.04}$		& 		17.42$^{+0.04}_{-0.03}$		&		16.08$^{+0.04}_{-0.02}$			&			16.1$\pm$0.1			&		6.76$\pm$0.01			&		6.24$\pm$0.01			\\
&		2010					&		16.48$\pm$0.03				& 		16.48$\pm$0.03				&		14.90$^{+0.03}_{-0.02}$			&			13.4$\pm$0.1			&		5.91$\pm$0.01			&		5.35$\pm$0.01			\\ \hline
&		Ratio (2010/2000)		&		0.929$^{+0.002}_{-0.003}$	& 		0.946$^{+0.003}_{-0.002}$	&		0.923$^{+0.003}_{-0.002}$		&			0.832$\pm$0.008			&		0.874$\pm$0.002			&		0.857$\pm$0.002			\\ \hline
\end{tabular}
\end{center}
\begin{scriptsize}
\tablenotetext{1}{The errors are at 1 $\sigma$ confidence level. }
\tablenotetext{2}{flux [${\times}10^{-11}$ erg cm$^{-2}$ s$^{-1}$]. }
\tablenotetext{3}{count rate [counts s$^{-1}$] }
\end{scriptsize}
\end{table*}

\section{Correction of proper motion effect}

\begin{figure}[h]
 \begin{center}
  \includegraphics[width=16cm]{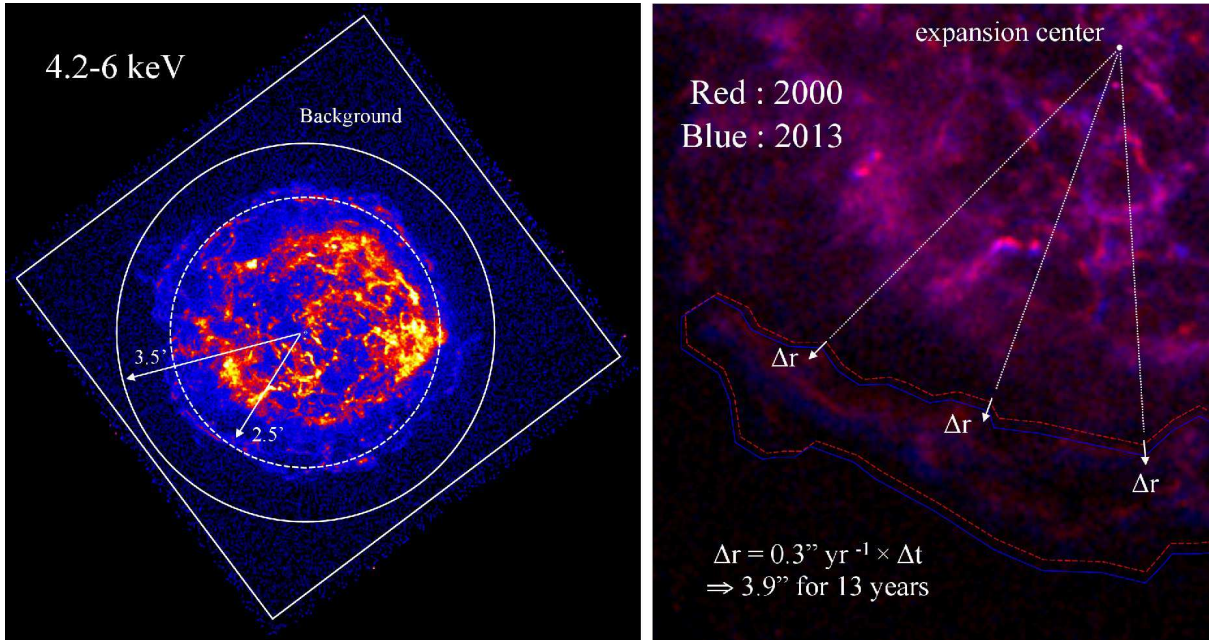}
 \end{center}
\caption{Left: difference of source region. The radius of the broken circle and solid circle are $r = 2.5^{\prime}$ and $r = 3.5^{\prime}$,
respectively. Right: region shift in the Forward Shock region. We shifted each apex of the polygon region to expansion direction
at $0.30^{''}$ \% yr$^{-1}$.}
\label{fig:region_def}
\end{figure}

The proper motion of the forward shock of Cassiopeia A has been well studied by X-ray \citep*{2007AJ....133..147P}, and its expansion rate
is $0.30^{''}$ \% yr$^{-1}$ on average. If we discuss a flux variation in the forward shock, we defined a region
which is shifting with the Forward Shock region at this rate. We adopt a polygon shape as the shape of the Forward Shock
region, and then we shifted each apex to expansion direction at $0.30^{''}$ \% yr$^{-1}$ (see Figure \ref{fig:region_def} right).
If we do not adapt this region shift, we found the flux in 2013 is $\sim5$ \% higher than that in 2000 because a component which was on the
exterior of the region 13 years ago leaked into the inside of the region. In the reverse shock region, there are less contribution
of leaks than in the forward shock since the proper motion of the reverse shock is small. When we adapt the region shift to the East
region, it is found that the decay rate does not change. Therefore, we have not adopt the region shift in the reverse shock regions.

\end{document}